**A combined high-pressure experimental and theoretical study of the electronic band-structure of scheelite-type AWO$_4$ (A = Ca, Sr, Ba, Pb) compounds**

R. Lacomba-Perales[1], D. Errandonea[1], A. Segura[1], J. Ruiz-Fuertes[1], P. Rodríguez-Hernández[2], S. Radescu[2], J. López-Solano[2], A. Mujica[2], and A. Muñoz[2]

[1] *Departamento de Física Aplicada-ICMUV, MALTA Consolider Team, Universitat de València, Edificio de Investigación, c/Dr. Moliner 50, 46100 Burjassot, Spain*

[2] *Departamento de Física Fundamental II, MALTA Consolider Team, Instituto de Materiales y Nanotecnología, Universidad de La Laguna, La Laguna, 38205 Tenerife, Spain*

## Abstract

The optical-absorption edge of single crystals of CaWO$_4$, SrWO$_4$, BaWO$_4$, and PbWO$_4$ has been measured under high pressure up to ~20 GPa at room temperature. From the measurements we have obtained the evolution of the band-gap energy with pressure. We found a low-pressure range (up to 7-10 GPa) where alkaline-earth tungstates present a very small E$_g$ pressure dependence (-2.1 < dE$_g$/dP < 8.9 meV/GPa). In contrast, in the same pressure range, PbWO$_4$ has a pressure coefficient of -62 meV/GPa. The high-pressure range is characterized in the four compounds by an abrupt decrease of E$_g$ followed by changes in dE$_g$/dP. The band-gap collapse is larger than 1.2 eV in BaWO$_4$. We also calculated the electronic-band structures and their pressure evolution. Calculations allow us to interpret experiments considering the different electronic configuration of divalent metals. Changes in the pressure evolution of E$_g$ are correlated with the occurrence of pressure-induced phase transitions. The band structures for the low- and high-pressure phases are also reported. No metallization of any of the compounds is detected in experiments nor is predicted by calculations.

## I. INTRODUCTION

Divalent-metal tungstates (AWO$_4$) with the tetragonal scheelite structure (S.G. *I4$_1$/a*) [1], are wide band-gap semiconductors [2], which present several technological applications [3-6] due to their excellent properties as scintillating crystals. The electronic structure of these compounds has been analyzed at ambient conditions [7,8]. Recently, accurate values of the band-gap energy (E$_g$) at ambient conditions have been reported [2]. Furthermore, E$_g$ has been revealed to be sensitive to the ionic radii and the electronic configuration of the divalent metal [2]. On the other hand, high pressure has been shown to be an excellent tool to study fundamental properties of these materials [9]. Actually, several high-pressure (HP) works studying their structural properties [10,11] and lattice dynamics [12-15] can be found in the literature. These works put on manifest the occurrence of pressure-induced phase transitions, which basically consist in a tetragonal to monoclinic symmetry reduction. However, there are very few studies concerning the behavior of the optical properties and the electronic structure of orthotungstates under HP [16-18]. In this paper we report the evolution of E$_g$ with pressure for CaWO$_4$, SrWO$_4$, BaWO$_4$, and PbWO$_4$. This information was obtained by means of HP optical-absorption measurements up to 20 GPa. On top of the experiments, we have also performed electronic band-structure calculations. Calculations were performed at different pressures within the framework of the density-functional theory (DFT) and considering the crystalline structure of the low- and high-pressure phases. Our combined experimental and theoretical study provides a clear picture of the changes induced by pressure in the electronic structure of scheelite-type tungstates.

## II. EXPERIMENTAL AND THEORETICAL DETAILS

Ultraviolet-visible (UV-VIS) transmittance measurements were carried out under compression at room temperature (RT) in a 500-μm culet diamond-anvil cell (DAC) [19]. From these experiments we determined the evolution with pressure of the optical-absorption edge of the studied tungstates. The absorption coefficient ($\alpha$) was obtained from the sample transmittance, thickness, and reflectivity. The pressure dependence of the thickness was taken into account [10, 11]. The refractive index is assumed to be constant [20]. In the set-up we used as light source a deuterium lamp of broad frequency spectrum. In order to avoid UV absorptions, fused silica lenses and Cassegrain reflectors (x15) were used. The absorption spectra were collected by a broad-range spectrophotometer (Ocean Optics USB4000-UV-VIS). Single crystals of $AWO_4$ (A = Ca, Sr, Ba and Pb) were grown by Czochralski method starting from raw powders of 5N purity [21-25]. We used thin platelets with no determined orientation (size: ~ 50x50x20 $\mu m^3$) directly cleaved from the bulk. A 200-$\mu m$ diameter hole drilled in a 40-$\mu m$ thick INCONEL gasket was used as pressure chamber. Methanol-ethanol-water (16:3:1) was the pressure-transmitting medium and pressure was determined using ruby photoluminescence [26]. Measurements were limited to 20 GPa because at higher pressures non reversible defects appear on the samples affecting their transmittance and not allowing an accurate determination of $\alpha$ and $E_g$

In the last years *ab initio* methods have allowed detailed studies of the crystal and electronic structure of materials under pressure [27]. In this work, band-structure calculations have been performed within the framework of DFT with the Vienna *ab initio* simulation package (VASP) [28]. Technical details of these calculations can be found in our previous works where the methodology is extensively described [10-11].

### III. RESULTS AND DISCUSSION

### A. Evolution of the band gap with pressure

The optical-absorption spectra of $CaWO_4$, $SrWO_4$, $BaWO_4$, and $PbWO_4$, obtained at several pressures from the studied samples, are shown in Fig. 1. The spectra measured at low pressures resemble those reported in literature at ambient conditions [2]. At 1 bar ($10^{-4}$ GPa), a steep absorption starts at 4.8, 4.9, 5.0, and 3.7 eV in $CaWO_4$, $SrWO_4$, $BaWO_4$, and $PbWO_4$, respectively. An absorption tail is also clearly seen at lower energies. This tail is typical of orthotungstates [2]. Its nature has been the subject of considerable debate and is beyond the scope of this work. This tail overlaps partially with the fundamental absorption but it does not preclude us to conclude that the fundamental band gap is direct; a conclusion that is also supported by our calculations with the exception of $PbWO_4$. The reason of this apparent discrepancy is that in $PbWO_4$ there is a direct and indirect gap separated by a few meV (see Table I). Therefore, in order to obtain quantitative information from our experiments, we will consider that in the four compounds the band gap is of the direct type and that the fundamental absorption edge obeys Urbach's rule [29, 30].

In Fig. 1, it can be seen that the band gap of $CaWO_4$ slightly decreases with pressure up to 9 GPa. At this pressure, which agrees with the structural transition pressure [9], there is an abrupt change of the pressure evolution of the absorption edge. The pressure dependence obtained for $E_g$ is given in Fig. 2 and its pressure coefficient ($dE_g/dP$) is shown in Table I. In $SrWO_4$ we observed a similar behavior in the low-pressure phase, but in this case the band gap slightly opens with pressure. This can be seen by comparing the spectra collected at 1 bar and 10.1 GPa. Beyond this pressure, the band gap reverts its pressure evolution recovering the ambient pressure value near 12.2 GPa (see figure). At this pressure there is an abrupt collapse of the band gap, after which the band gap red-shift at -81(6) meV/GPa (see Fig. 2). $BaWO_4$ shows a behavior qualitatively similar to that of $SrWO_4$, but in $BaWO_4$ the band-gap collapse takes place at 7.4 GPa and it is much larger than in $SrWO_4$ ($\Delta E_g > 1.2$ eV). Finally, $PbWO_4$ is the only compound that shows a different behavior for the low-pressure phase. In this compound, in the scheelite structure the band gap red-shifts at -62(2) meV/GPa in good agreement with previous published experiments [16]. Near the pressure-induced phase transition there is also a band-gap collapse. The reason for the changes produced for pressure in the absorption edge will be discussed in the following sections in relation with the structural stability and the band structure of different phases. To conclude this section, we would like to add that changes induced by pressure in the absorption edge are reversible in the four studied compound within the pressure range covered by these experiments.

**B. Low-pressure range**

In Fig. 3 band dispersions for the four materials are plotted along different symmetry directions within the body-centered tetragonal Brillouin zone of the scheelite structure. Brillouin zones of scheelite and other structures relevant for this work are shown in Fig. 4. The shapes of the bands for $CaWO_4$, $SrWO_4$, and $BaWO_4$ are very similar to each other. In $CaWO_4$, $SrWO_4$ the valence-band maxima and conduction-band minima are located at the $\Gamma$ point, so that these are direct-gap materials as commented in the previous section. In $BaWO_4$ we found two band gaps very close in energy: a direct band-gap ($\Gamma$-$\Gamma$) with $E_g = 4.624$ eV and an indirect-one ($\Gamma$-$Z$) with $E_g = 4.617$ eV. However, these differences are comparable to the error of calculations and since experiments show that $BaWO_4$ is a direct-gap material and the absorption intensity of indirect transitions is expected to be much lower, we will use the first value to compare with experiments. Although the values of the band gaps calculated within density-functional theory are known to be underestimated, it is interesting to compare the gaps calculated for the three materials as listed in Table I. In the three compounds $E_g$ is underestimated by the calculations at least by 0.6 eV. This underestimation increases following the sequence $BaWO_4 < SrWO_4 < CaWO_4$. Similar results have been obtained for $CaWO_4$ and $BaWO_4$ from previous calculations [7, 31].

One important feature to remark of the band structure of the scheelite phase in the alkaline-earth tungstates is that the dispersion of the valence bands is relatively small, with comparable dispersions along different directions. In addition, the upper part of the valence band is dominated by O *2p* states. On the other hand, the lower part of the conduction band, which is composed primarily of electronic states associated with the W *5d* states, is separated by approximately 0.5 eV from the upper part of the conduction band formed from states of W and the *3d, 4d, 5d* states of Ca, Sr, and Ba, respectively. Fig. 5 shows the calculated density of states for scheelite illustrating the contribution of different states.

The shape of the bands for $PbWO_4$ is somewhat different from those of the other materials. For $PbWO_4$, the band extrema are located away from the $\Gamma$ point. Within the region of the Brillouin zone studied for the dispersion plot shown in Fig. 3, we can state that valence band has maxima in the $\Delta$ directions and the conduction-band minima are located in the $\Sigma$ directions. In addition, very close to the absolute maxima of the valence band there are maxima at the $\Sigma$ direction. Then according with calculations $PbWO_4$ has an indirect gap of 3.19 eV and a direct gap of 3.27 eV. The proximity between both gaps could have made the indirect band gap undetectable in experiments, as previously commented for $BaWO_4$. The band structure obtained from the calculations is in good agreement with that reported by Zhang *et al* [7]. There are only differences smaller than 10% in the value of the band gaps. In particular, the present calculations give band-gap energies that differ less from the experiments. As in the case of the alkaline-earth tungstates, calculations show that in $PbWO_4$, the valence and conduction bands are mainly composed of O *2p* states and W *5d* states. However, Pb *6s* states to some extent also contribute to the top of the valence band and bottom of the conduction band, being a distinctive feature of $PbWO_4$.

The calculated ordering of the band gaps at ambient pressure is given by $PbWO_4$ < $CaWO_4$ < $SrWO_4$ < $BaWO_4$. Experimental data agree with this sequence. $PbWO_4$ has the smallest gap given the particular electronic configuration of Pb. In alkaline-earth tungstates only the *s* orbitals of the divalent cation make a minor contribution to the valence and conduction bands. On the other hand, in $PbWO_4$ there are two Pb *s* electrons in the valence band, and the $O^{2-}$ *2p* states and the $W^{6+}$ *5d* also hybridize with the *s* and *d* states of $Pb^{2+}$. From symmetry considerations (inversion symmetry), the Pb *6s* and O *2p* states are expected not to mix in the $\Gamma$ point but to strongly mix in directions with lower symmetry. The resulting *s-p* repulsion pushes up (down) the maximum (minimum) of the valence (conduction) band resulting in a band gap away from the $\Gamma$ point. These conclusions have been confirmed by x-ray photoelectron spectroscopy measurements performed in $BaWO_4$, $CaWO_4$, and $PbWO_4$ [8].

Regarding the pressure evolution of the band gap, calculations give a similar behavior than experiments for $CaWO_4$ and $PbWO_4$. The obtained pressure evolution for all compounds is shown in Fig. 6 and pressure coefficients are given in Table I. Some discrepancies are found for $SrWO_4$. For this compound the calculations predict that the gap closes with pressure ($dE_g/dP$ = -4.3 meV/GPa) but from the experiments the opposite behavior is obtained ($dE_g/dP$ = + 3.7 meV/GPa). In the case of $BaWO_4$, calculations predict the direct gap slightly to open under compression, but the indirect gap to close. In the experiments the gap slightly opens under compression. The origin of the discrepancies for $SrWO_4$ is not clear yet. One possible explanation is the existence of excitonic effects that are not taken into account in DFT calculations but are normally strong in the absorption edge of direct semiconductors. Small subtle changes in the band structure under pressure (like a close indirect transition becoming closer or further in energy) can strongly change the exciton life and, consequently, its width. A decrease of the exciton width would lead to a steeper absorption tail (as it seems to be the case for SrWO4 in Fig. 1) and compensate the band gap red shift, as was observed for the direct gap of $CuAlO_2$ [32]. Measurements with thinner samples would be necessary to elucidate the origin of this small discrepancy. However, differences on pressure coefficients are of the order of a few meV/GPa. Such differences are comparable with the error on gap determination both in experiments and calculations. Another possibility to explain discrepancies can be differences in temperature and hydrostaticity between experiments and calculations.

By comparing the density of states at different pressures it is possible to explain the changes induced by pressure in the band-structure of scheelites. In order to do it, the density of states is plotted at 5.5 GPa for the four different compounds in Fig. 5(b). There it can be seen that most notorious changes in the density of states are induced in $PbWO_4$ and $BaWO_4$. According with the calculations, the reduction of the band gap in $CaWO_4$ is a consequence of the increase of the contribution of Ca *3d* states to the valence band. A similar effect, but most moderate is predicted for $SrWO_4$. According to this, pressure should produce a reduction of $E_g$, but we observe experimentally the opposite as discussed above. Regarding $PbWO_4$, its distinctive band structure, where Pb states play a more important role than Ca, Sr, and Ba states in other compounds, makes the effect of pressure to be more prominent, being $dE_g/dP$ an order of magnitude larger than in other compounds. As a consequence of it the electronic band-gap is reduced under compression more in $PbWO_4$ than in any of the other three compounds. This can be clearly seen by comparing Figs. 5(a) and 5(b). This conclusion is consistent with the fact that in scheelite-type $PbMoO_4$ a $dE_g/dP$ similar to -62 meV/GPa was determined [17]. Note that according to band-structure calculations $PbWO_4$ and $PbMoO_4$ have a quite similar electronic structure [7]. A similar band-structure is expected for $EuWO_4$ [33], in which pressure should induce a similar closing of the band-gap. On the other hand, $CaMoO_4$ has a similar band structure to $CaWO_4$ [7], which suggests that pressure should have little effects on the band structure of $CaMoO_4$ within the stability range of the scheelite phase.

**C. High-pressure range**

Several phase transitions are induced by pressure in the studied compounds. In particular in $CaWO_4$ and $SrWO_4$ a phase transition to the monoclinic fergusonite (*I2/a*) structure is known to occur near 10 GPa [10]. On the other hand, in $BaWO_4$ and $PbWO_4$ not only the fergusonite structure appears under compression but also another monoclinic phase is observed, with *P2₁/n* symmetry named $BaWO_4$-II (or $PbWO_4$-III) [11]. One or the other structure has been found to be the post-scheelite phase depending upon experimental conditions and in some cases both phases coexist in a broad pressure range. In our case we will show that the scheelite-*P2₁/n* transition explains better the experimental changes observed in the optical absorption [34].

Let us discuss first $CaWO_4$. In this compound our calculations found a collapse smaller than 0.1 eV at the scheelite-fergusonite transition. This is in good agreement with the experiments. We also found that the band gap of fergusonite is direct and located at the $\Gamma$ point (see Fig. 3). The band structure of fergusonite resembles closely that of scheelite, which is not surprising because fergusonite is a monoclinic distortion of scheelite. In addition, around 12 GPa the calculations predict a direct-to-indirect band crossing in the HP phase, changing the lowest minimum of the conduction band to the Y point of the Brillouin zone. This minimum moves much faster with pressure than the minimum at the $\Gamma$ point and therefore the pressure coefficient becomes -105 meV/GPa (see Fig. 6). This picture fully agrees with the results found in the experiments as can be seen in Fig. 2. A similar behavior is found for $SrWO_4$, but in this case the band gap collapse is larger than in $CaWO_4$.

In the cases of $BaWO_4$ and $PbWO_4$ a monoclinic $P2_1/n$ structure fits better than a fergusonite with the changes observed in the band gap after the phase transition. The band structure of the $P2_1/n$ high-pressure structure is different than that of fergusonite. In particular, in the $P2_1/n$ phase the band-gap is much smaller than in fergusonite, explaining the larger collapse of the gap at the transition. We also found that the high-pressure phase of $PbWO_4$ is a direct band-gap semiconductor with the fundamental gap at the D point of the Brillouin zone. On the other hand, in the high-pressure phase of $BaWO_4$ there are two gaps, one direct (D - D) and one indirect ($\Gamma$–D) very close in energy. Regarding the pressure coefficient of the gap, the calculated values explain the experimental results. In the $P2_1/n$ structure the effect of pressure is smaller than in fergusonite. This is consistent with the fact that $BaWO_4$-II and $PbWO_4$-III are quite dense structures with a low compressibility.

We would like to comment on differences between the experimental results reported here for the HP phase of $PbWO_4$ and those previously published [16]. In contrast with the previous study we only found one abrupt change in $E_g$ and supported by calculations we attribute it to the scheelite-to-$PbWO_4$-III transition. Previously, a collapse of $E_g$ and a second change were detected. These differences are not surprising since a rich polymorphism is induced by pressure in $PbWO_4$ and $BaWO_4$ [34]. In addition, we think that differences can be caused by the different thickness of the samples used. In this case we used a thinner sample than in Ref. [16]. This could have lead to better quasi-hydrostatic conditions avoiding therefore the appearance of the fergusonite phase in between scheelite and $PbWO_4$-III and making the formation of defects to appear at higher pressures. The improvement in the experimental conditions in the present experiments has also leaded to a better agreement between experiments and theory.

In the past, it has been argued that metallization can be induced by pressure in scheelite-structured oxides due to the increase of electronic hybridization [35]. In our case both experiments and calculations do not point towards pressure-induced metallization. In the case of $PbWO_4$, the compound with the smallest gap, the band gap raches 2.3 eV at 18 GPa. This is consistent with color changes observed in the crystals, which become greenish when approaching this pressure. Upon further compression, we were unable to determine $E_g$ because of the appearance of many defects in the crystals. However, we have observed that around 25 GPa the crystals became orange, what indicates that the band-gap cannot be smaller than 2 eV. Calculations predict a similar behavior for $E_g$ in the HP phase. Extrapolating $E_g$ to higher pressures, using the experimental pressure coefficient, metallization is estimated to take place at 100 GPa. However, at much lower pressures additional phase transitions are predicted to take place [10, 11] and amorphization has been observed [36].

## IV. CONCLUSIONS

Absorption spectra of $CaWO_4$, $SrWO_4$, $BaWO_4$, and $PbWO_4$ were measured as a function of pressure up to 20 GPa extending the pressure range of previous experiments that were only performed for lead tungstate. In the low-pressure scheelite phase we observed in $PbWO_4$ a red-shift of the absorption edge under compression. In the other three compounds the band gap is much less affected by pressure. At the transition pressure to different monoclinic phases an abrupt change of the absorption spectrum was found. Additionally, in $CaWO_4$ and $SrWO_4$ a band crossover occurs in the HP phase. The changes of the optical-absorption edge were attributed to the occurrence of previously observed phase transitions. The results are explained by means of high-pressure electronic structure calculations for the different structures of the four studied compounds.

**Acknowledgements**

Research financed by the Spanish MEC under Grants No. MAT2010-21270-C01/03 and No. CSD-2007-00045. We acknowledge the supercomputer time provided by the Red Española de Supercomputación (RES).

**Table I.** Experimental and theoretical values of the band-gaps at ambient pressure, $E_g(0)$ in eV, and pressure coefficients, $dE_g/dP$ in meV/GPa. Values are obtained from linear fits at the different pressure ranges from Fig. 2 – experiment - and Fig. 6 – theory. We also listed the k-points for the top of the valence band and bottom of the conduction band. Pressure is given in GPa.

| CaWO₄ | Experiment | | Theory | | |
|---|---|---|---|---|---|
| | | | Scheelite | Fergusonite | |
| **P** | 0.2 − 9.1 | 9.1 − 13.4 | 0 − 7.5 | 9.4 − 12.9 | 13.1 − 18.5 |
| **$E_g(0)$** | 4.94 ± 0.02 | 5.57 ± 0.04 | $(\Gamma \to \Gamma)$ 3.84 | $(\Gamma \to \Gamma)$ 3.76 | $(\Gamma \to Y)$ 4.63 |
| **$dE_g/dP$** | -2.1 ± 0.3 | -73 ± 3 | -8.4 | -4.7 | -105 |

| SrWO₄ | Experiment | | | Theory | | |
|---|---|---|---|---|---|---|
| | | | | Scheelite | Fergusonite | |
| **P** | 0 − 10.1 | 10.1 − 12.2 | 12.5 − 15.5 | 0 − 8.3 | 9.8 − 14 | 14 − 20.3 |
| **$E_g(0)$** | 4.98 ± 0.04 | 5.06 ± 0.02 | 5.68 ± 0.08 | $(\Gamma \to \Gamma)$ 4.20 | $(\Gamma \to \Gamma)$ 4.34 | $(\Gamma \to Y)$ 5.05 |
| **$dE_g/dP$** | 3.7 ± 0.7 | -8 ± 2 | -81 ± 6 | -4.3 | -18.9 | -70.5 |

| BaWO₄ | Experiment | | Theory | | | |
|---|---|---|---|---|---|---|
| | | | Scheelite | | BaWO₄-II | |
| **P** | 0 − 6.4 | 7.4 − 15.9 | 0 − 7.5 | | 9.4 − 14.5 | |
| **$E_g(0)$** | 5.20 ± 0.03 | 4.10 ± 0.05 | $(\Gamma \to \Gamma)$ 4.624 | $(\Gamma -Z)$ 4.617 | $(\Gamma \to D)$ 3.45 | $(D \to D)$ 3.54 |
| **$dE_g/dP$** | 8.9 ± 0.8 | -11 ± 5 | 3.9 | -3.1 | -8.7 | -6.8 |

| PbWO₄ | Experiment | | Theory | | |
|---|---|---|---|---|---|
| | | | Scheelite | | PbWO₄-III |
| **P** | 0 − 6.4 | 7.4 − 19.4 | 0 − 5.2 | | 6.6 − 17.9 |
| **$E_g(0)$** | 4.01 ± 0.02 | 3.13 ± 0.02 | $(\Delta \to \Sigma)$ 3.19 | $(\Sigma \to \Sigma)$ 3.27 | $(D \to D)$ 2.74 |
| **$dE_g/dP$** | -62 ± 2 | -30 ± 2 | -63.9 | -54.1 | -22.3 |

**Figure Captions**

**Figure 1:** Selected optical-absorption spectra for different pressures in each of the materials studied.

**Figure 2:** Experimental evolution of the band gap with pressure. Solid points represent experimental data collected upon pressure increase, while open points correspond to experimental data collected upon pressure release. The solid lines are the fitting curves and the vertical dotted lines represent transition pressures. The different optical transitions related to the fundamental gap are also indicated according with the interpretation of theoretical calculations.

**Figure 3:** Electronic band-structure dispersion curves. (a) for $CaWO_4$, $SrWO_4$, $PbWO_4$, and $BaWO_4$ in the scheelite structure at ambient pressure. (b) for fergusonite CaWO4 and SrWO4 at 12.9 and 14 GPa, respectively, for $BaWO_4$-II at 11.5 GPa, and for $PbWO_4$-III at 12.8 GPa.

**Figure 4:** Diagram of the Brillouin zone for the scheelite, fergusonite, and monoclinic $P2_1/n$ structures.

**Figure 5:** (color online) Density of states of the schelite phase at ambient pressure (a) and 5.5 GPa (b).

**Figure 6:** Calculated evolution of the band gap with pressure for the stable phase in each pressure range. The lines are fitting curves and the symbols represent the points of the Brillouin zone where the optical transition takes place. For the fergusonite phases of $CaWO_4$ and $SrWO_4$, the evolution of the $\Gamma \rightarrow \Gamma$ and the $\Gamma \rightarrow Y$ gaps is shown to remark the band-crossing present in these materials. For the scheelite phase of $BaWO_4$ and $PbWO_4$ we show the direct and indirect gaps.

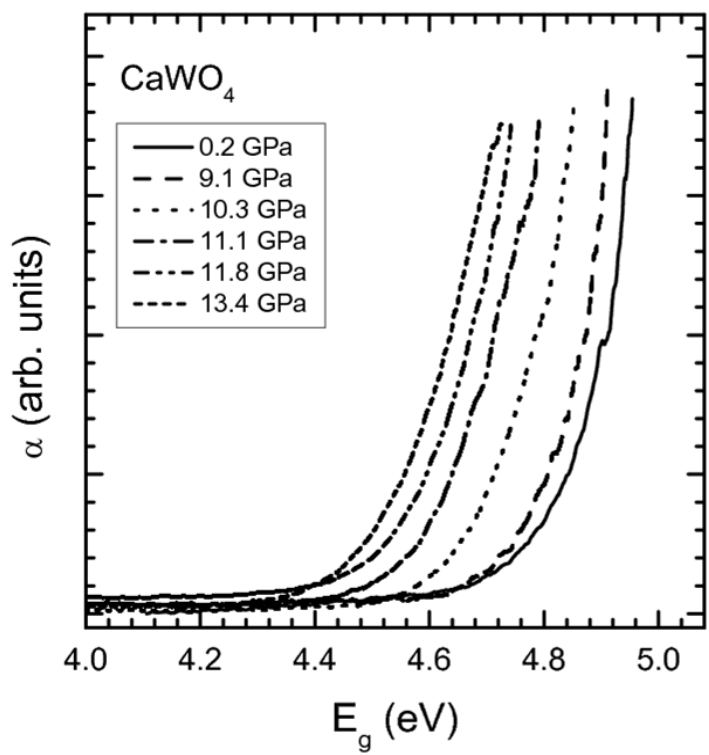

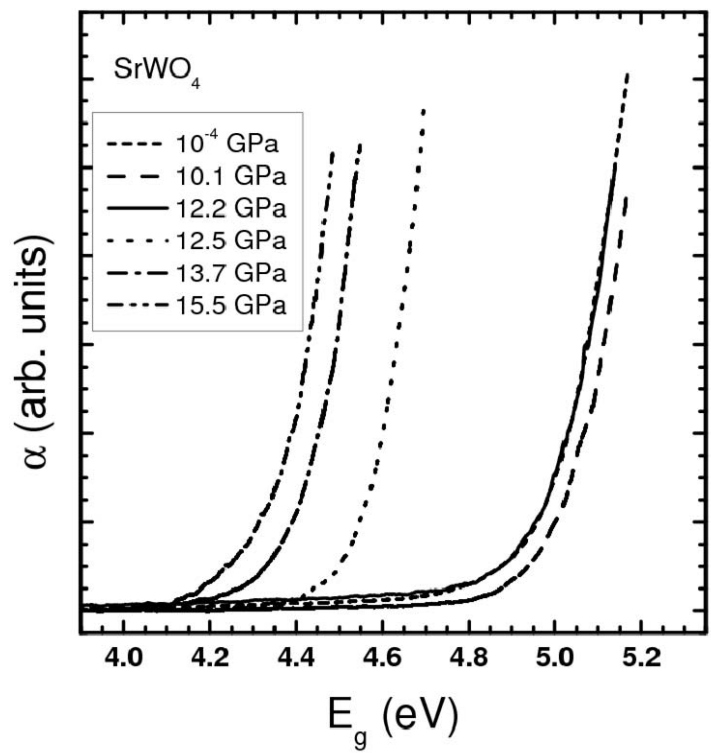

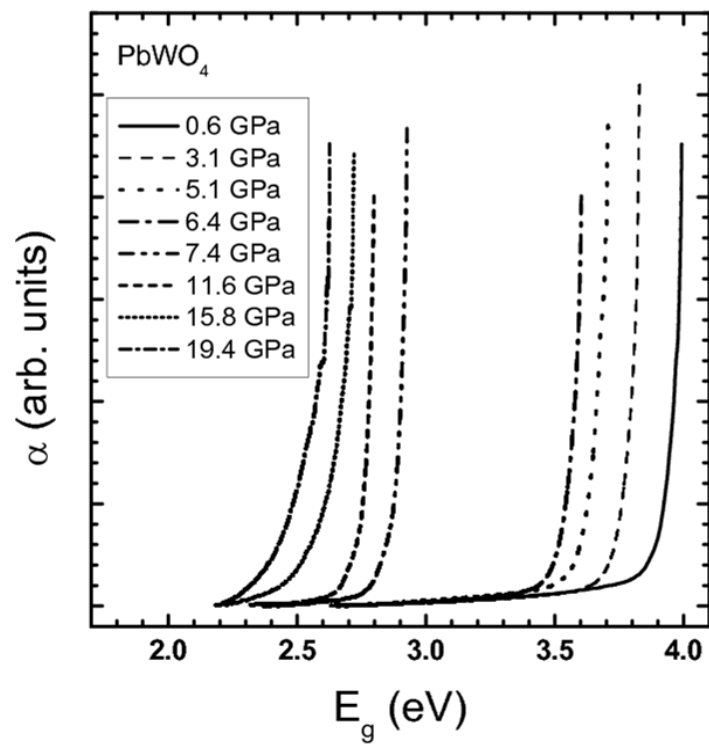

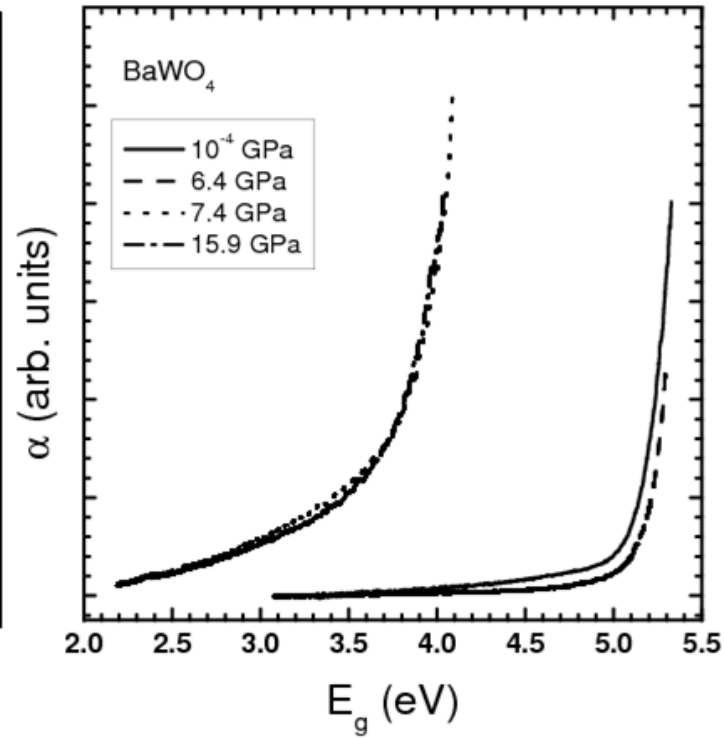

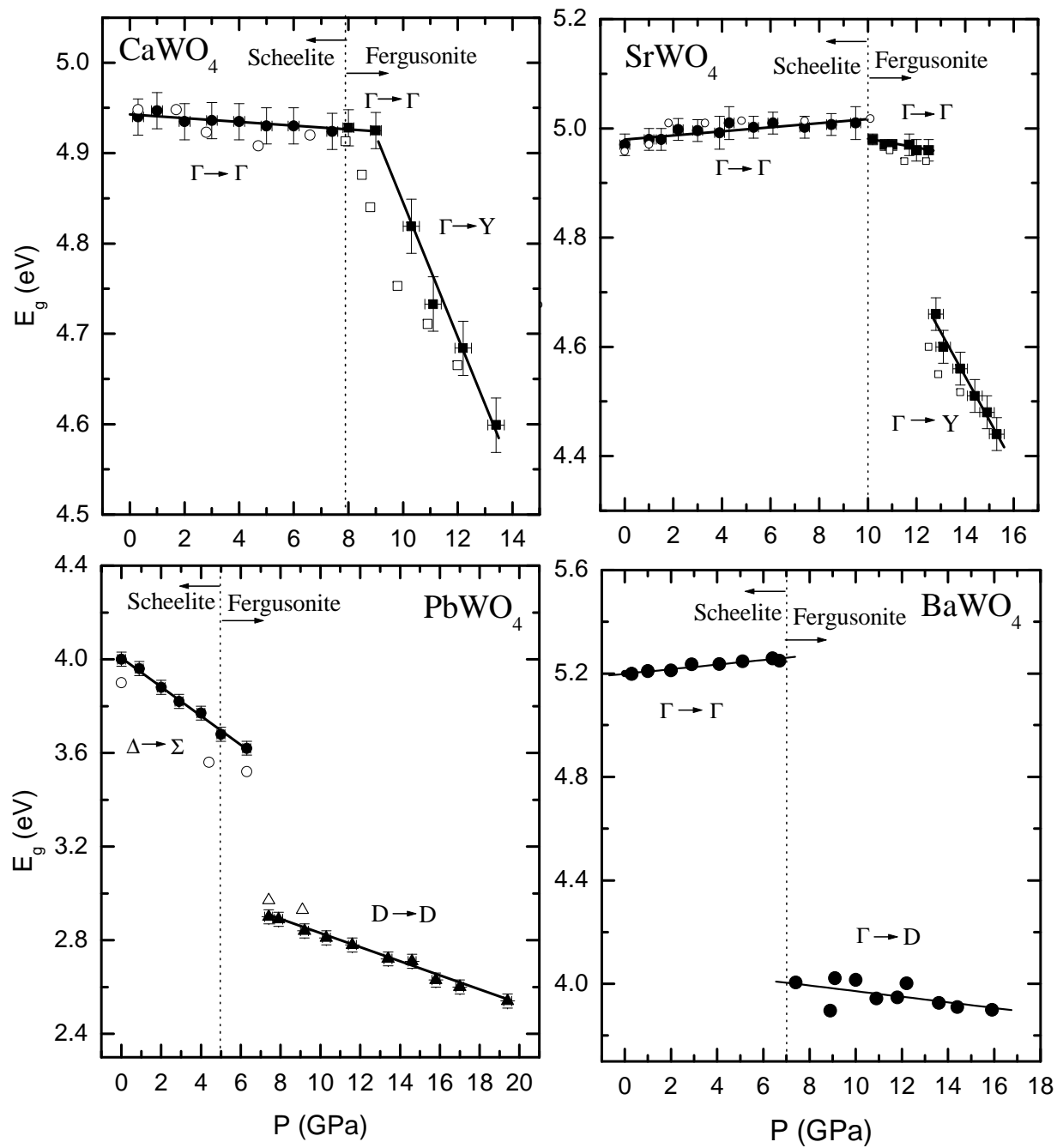

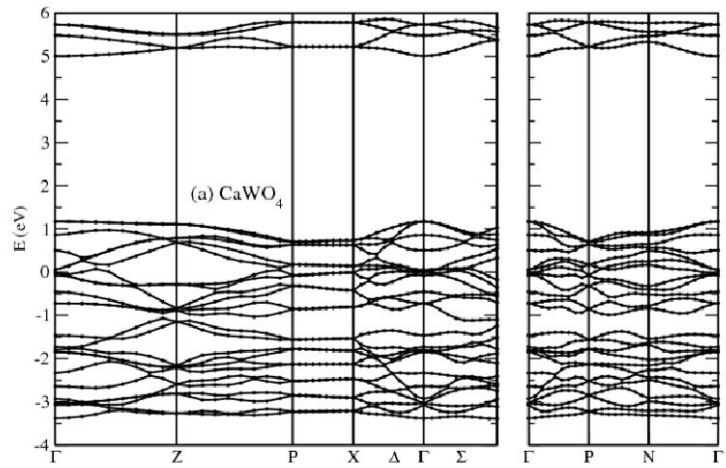

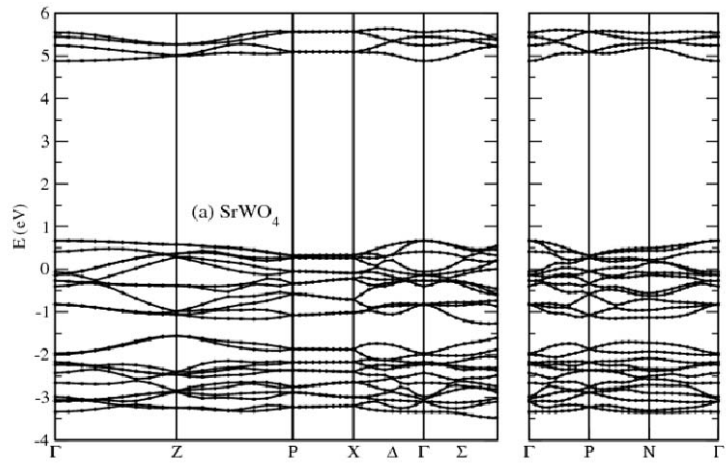

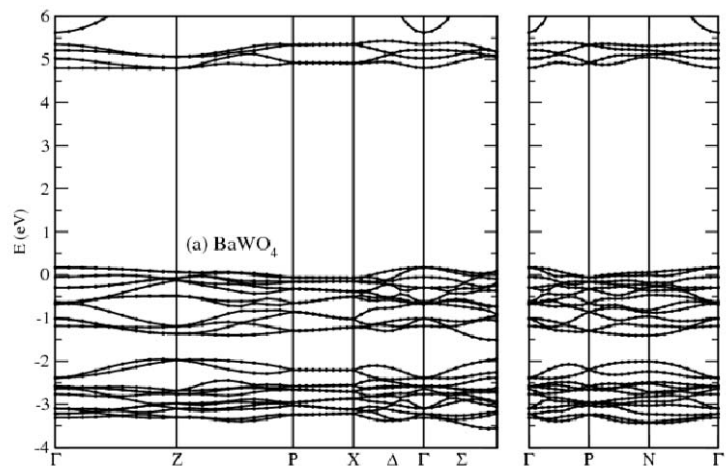

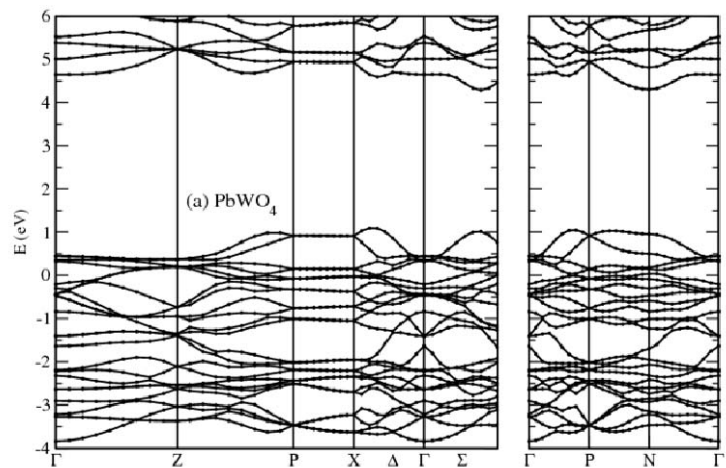

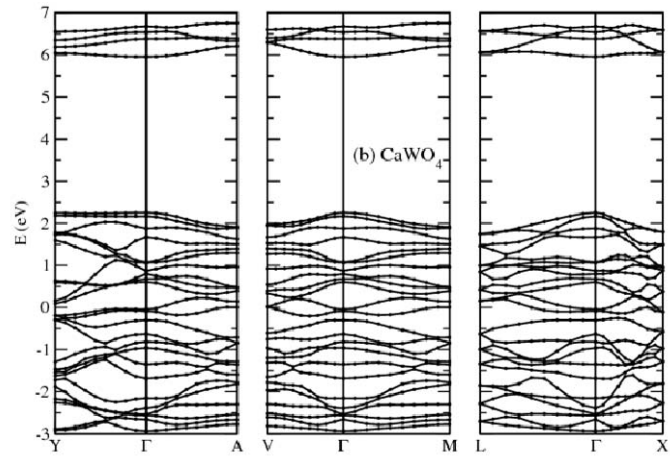

(b) CaWO$_4$

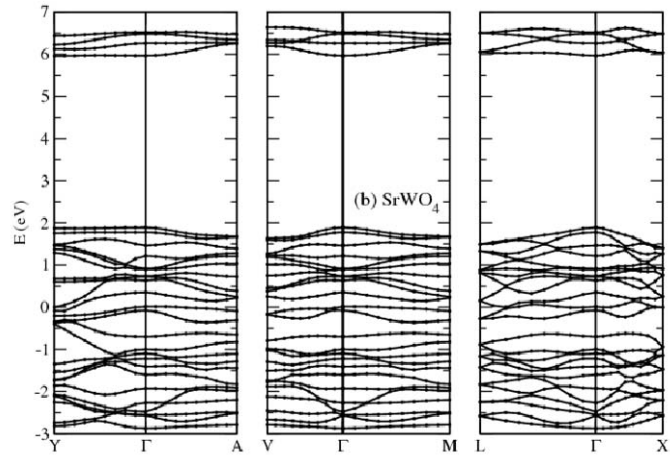

(b) SrWO$_4$

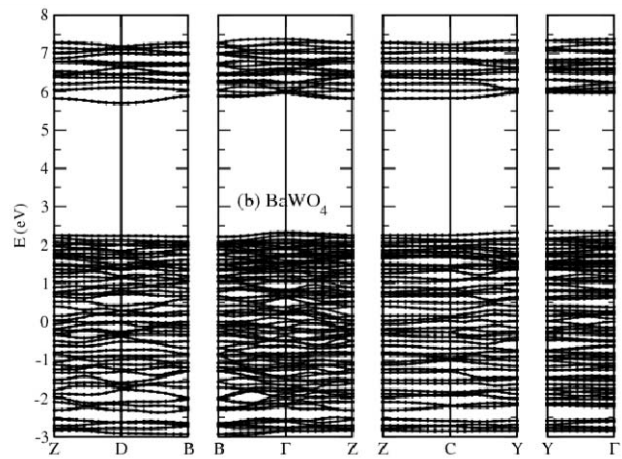

(b) BaWO$_4$

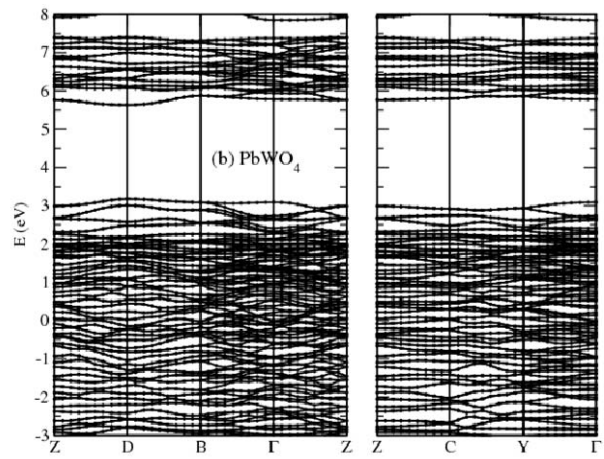

(b) PbWO$_4$

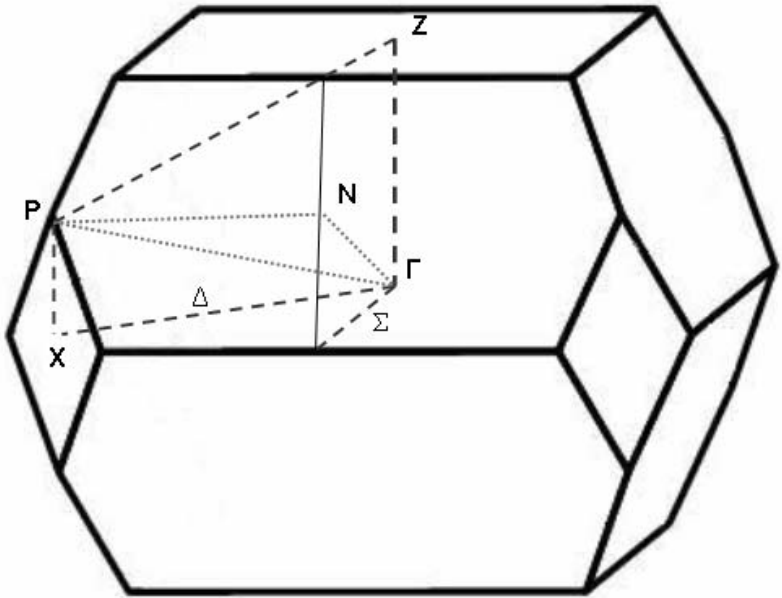

scheelite

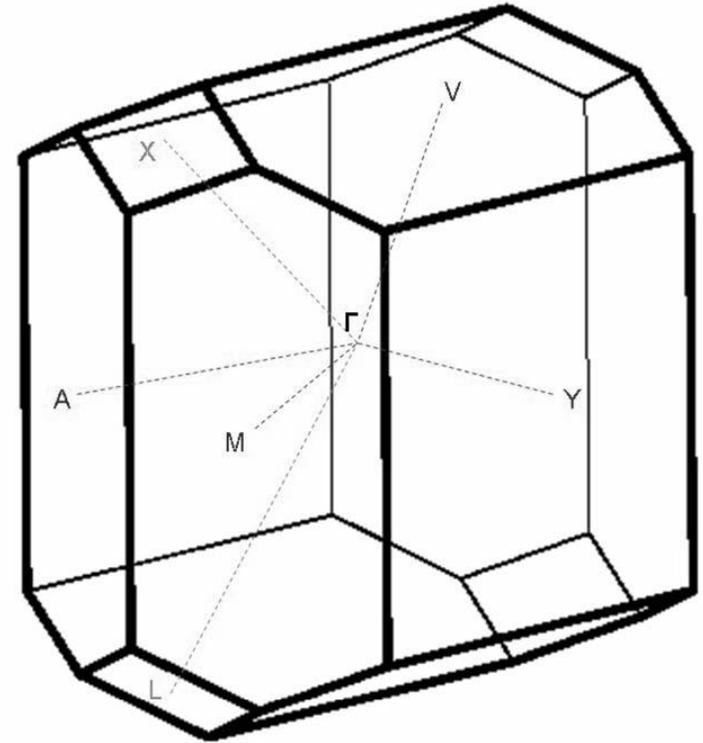

fergusonite

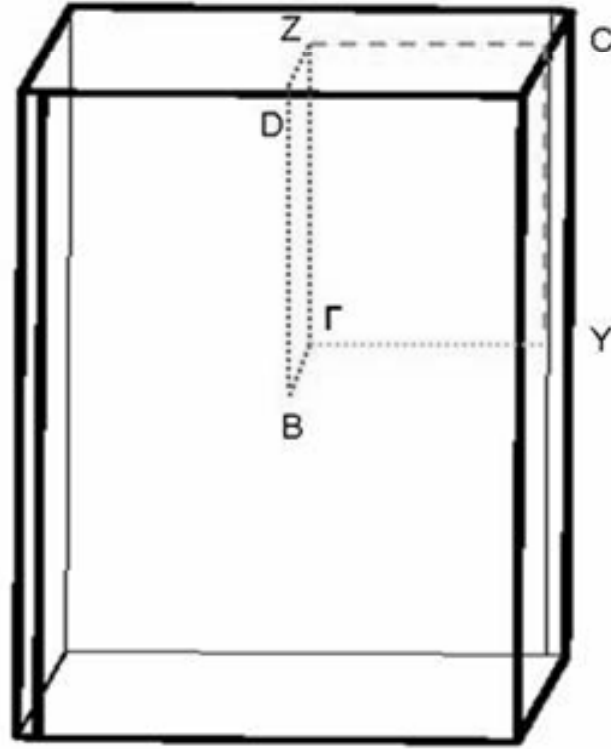

BaWO$_4$-II = PbWO$_4$-III

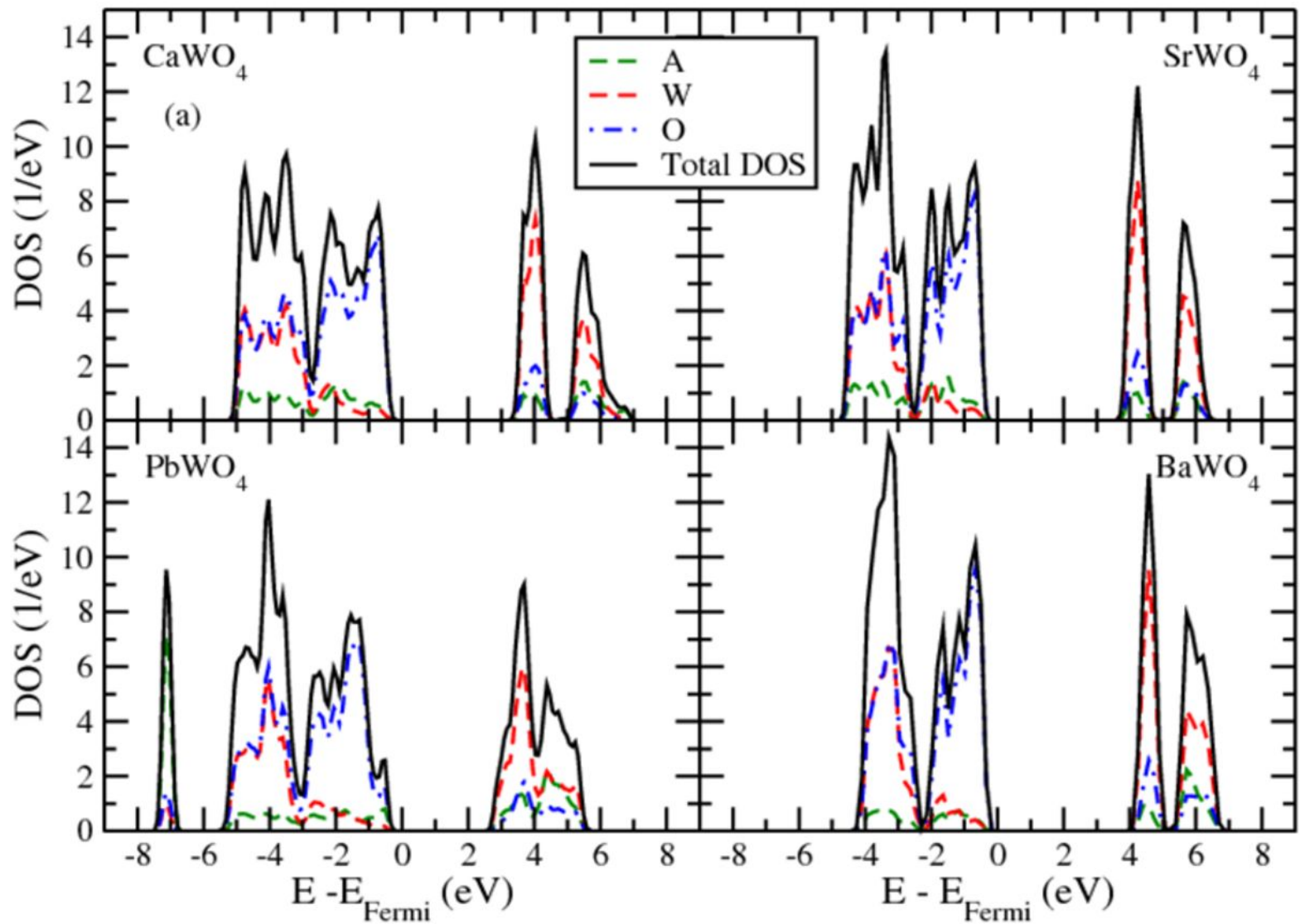

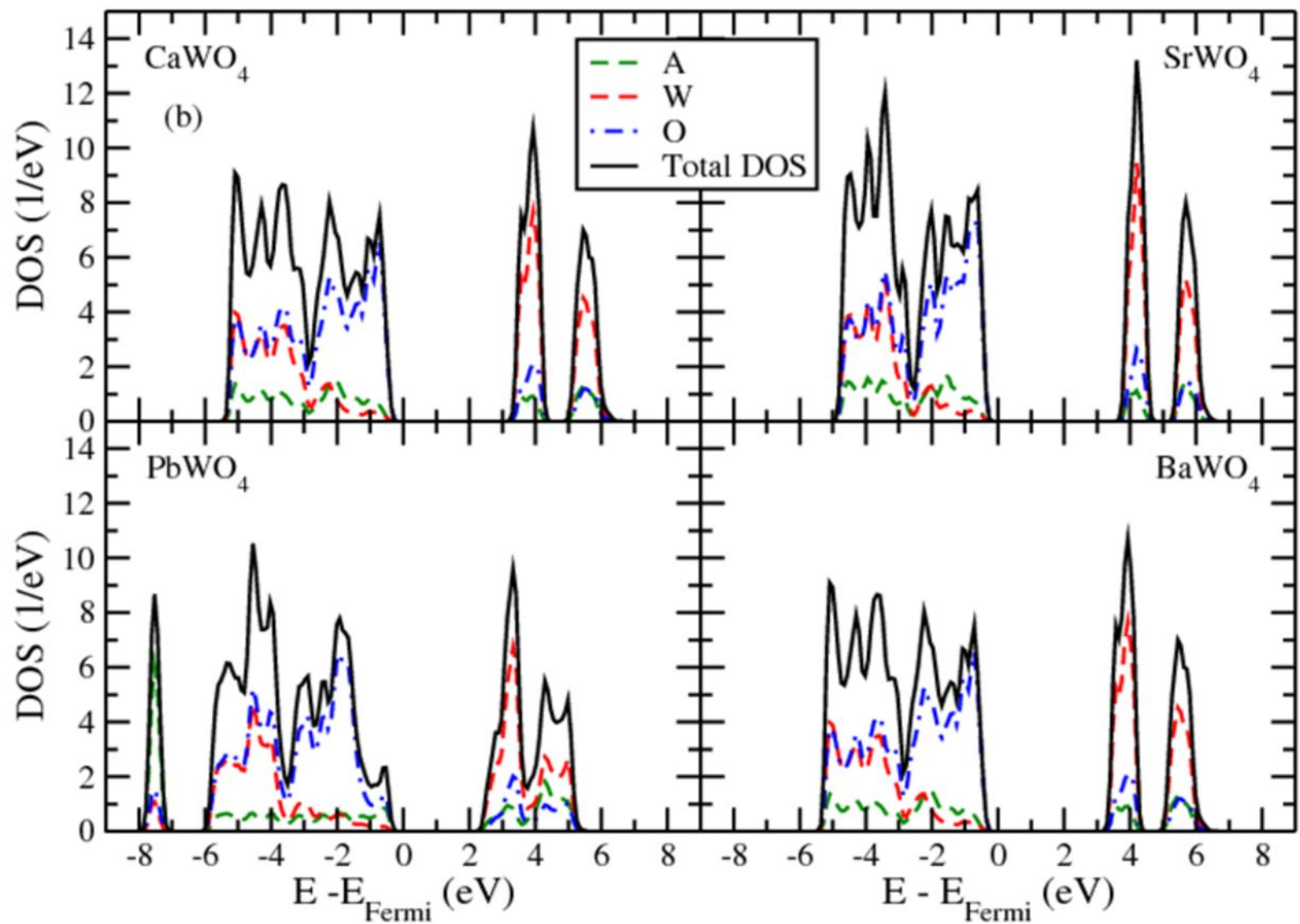

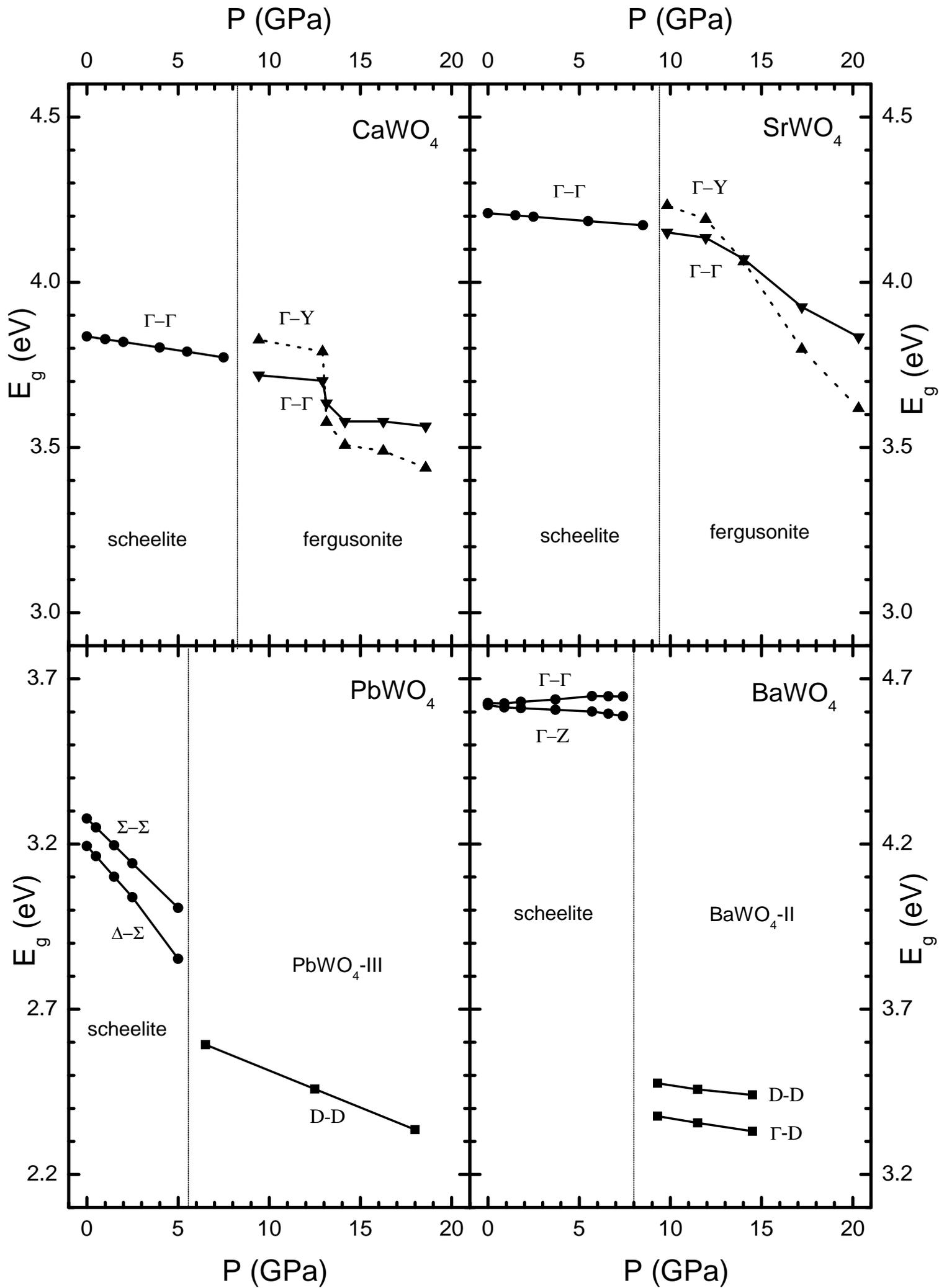